\documentclass[12pt]{article}

\usepackage{amsmath,amssymb,graphicx} %use drftcite in draftmode
\usepackage[usenames]{color}
\usepackage{multicol}
\usepackage{cite}

\definecolor{verdes}{cmyk}{0.92,0,0.59,0.4} 
\ifx\pdfoutput\undefined
\usepackage[dvips,bookmarks]{hyperref}	% This is for arXiv.org
\else
\usepackage{hyperref}	% This is for pdftex
\fi
\hypersetup{colorlinks,bookmarksopen,bookmarksnumbered,citecolor=verdes,
linkcolor=blus,pdfstartview=FitH,urlcolor=rossos}
\newcommand{\SU}{{\rm SU}}

\font\tenrsfs=rsfs10 at 12pt
\font\sevenrsfs=rsfs7
\font\fiversfs=rsfs5
\newfam\rsfsfam
\textfont\rsfsfam=\tenrsfs
\scriptfont\rsfsfam=\sevenrsfs
\scriptscriptfont\rsfsfam=\fiversfs
\def\mathscr#1{{\fam\rsfsfam\relax#1}}
\def\Lag{\mathscr{L}}

\newcommand{\TeV}{\,{\rm TeV}}

\newcommand{\NP}{Nucl. Phys.}

\newcommand{\CL}{\,\hbox{\rm CL}}

\makeatletter
\def\art{\@ifnextchar[{\eart}{\oart}}
\def\eart[#1]#2#3#4#5#6{{\rm #2}, {#3 \rm #4} {\rm (#6) #5 [#1]}}
\def\hepart[#1]#2{{\rm #2, #1}}
\newcommand{\oart}[5]{{\rm #1}, {\em #2 \rm #3} {\rm (#5) #4}}

\newcounter{alphaequation}[equation]
\def\thealphaequation{\theequation\hbox to
0.6em{\hfil\alph{alphaequation}\hfil}}
\def\eqnsystem#1{
\def\@eqnnum{{\rm (\thealphaequation)}}
\def\@@eqncr{\let\@tempa\relax \ifcase\@eqcnt \def\@tempa{& & &} \or
  \def\@tempa{& &}\or \def\@tempa{&}\fi\@tempa
  \if@eqnsw\@eqnnum\refstepcounter{alphaequation}\fi
\global\@eqnswtrue\global\@eqcnt=0\cr}
\refstepcounter{equation} \let\@currentlabel\theequation \def\@tempb{#1}
\ifx\@tempb\empty\else\label{#1}\fi
\refstepcounter{alphaequation}
\let\@currentlabel\thealphaequation
\global\@eqnswtrue\global\@eqcnt=0 \tabskip\@centering\let\\=\@eqncr
$$\halign to \displaywidth\bgroup \@eqnsel\hskip\@centering
$\displaystyle\tabskip\z@{##}$&\global\@eqcnt\@ne
\hskip2\arraycolsep\hfil${##}$\hfil& \global\@eqcnt\tw@\hskip2\arraycolsep
$\displaystyle\tabskip\z@{##}$\hfil
\tabskip\@centering&\llap{##}\tabskip\z@\cr}
\def\endeqnsystem{\@@eqncr\egroup$$\global\@ignoretrue} \makeatother

\newcommand{\beq}{\begin{eqnarray}}% can be used as {equation} or  {eqnarray}
\newcommand{\eeq}{\end{eqnarray}}

\newcommand{\centeron}[2]{{\setbox0=\hbox{#1}\setbox1=\hbox{#2}\ifdim

\wd1>\wd0\kern.5\wd1\kern-.5\wd0\fi
\copy0

\kern-.5\wd0\kern-.5\wd1\copy1\ifdim\wd0>\wd1
                                       \kern.5\wd0\kern-.5\wd1\fi}}
\newcommand{\ltap}{\>\centeron{\raise.35ex\hbox{$<$}}
                               {\lower.65ex\hbox{$\sim$}}\>}
\newcommand{\gtap}{\>\centeron{\raise.35ex\hbox{$>$}}
                               {\lower.65ex\hbox{$\sim$}}\>}

\newcommand\ZZ{\hbox{\zfont Z\kern-.4emZ}}
\font\zfont = cmss10 %scaled \magstep1

\renewcommand{\theequation}{\thesection.\arabic{equation}}
\definecolor{blus}{cmyk}{1,1,0,0.6}
\definecolor{rossos}{cmyk}{0,1,1,0.55}

\begin{document}
%\begin{titlepage}
%\begin{flushright}
%{\tt hep-ph/yymmnn}
%\end{flushright}
\phantom{X}
\vspace{1cm}
\begin{center}

{\huge\color{rossos} \bf   The Minimal Set of\\[3mm]
 Electroweak Precision Parameters}
%\\
%for Model Building}
%Ideas for titles:
% A Simple analysis of electroweak precision data
% A simple general method for analyzing electroweak precision constraints
% The minimal set of electroweak precision parameters
% An efficient method for electroweak precision constraints
% An efficient analysis of electroweak precision constraints
% An efficient parametrization of electroweak precision constraints

% Unveiling electroweak precision constraints: an efficient parametrization of non-universal new physics

\vspace{1cm}\bigskip

{
{\large\bf G. Cacciapaglia$^a$, C. Cs\'aki$^a$, G. Marandella$^b$,  and A. Strumia$^{c}$}
}

\bigskip\bigskip

{\it $^a$ Institute for High Energy Phenomenology \\ Newman Laboratory of Elementary Particle Physics \\ Cornell University, Ithaca, NY, 14853, USA}\\[3mm]
{\it $^b$ Department of Physics, University of California, Davis, CA 95616, USA}\\[3mm]
{\it $^c$ Dipartimento di Fisica dell'Universit{\`a} di Pisa and INFN, Italia}\\[1cm]

\vspace{1cm}

{\bf\large Abstract}
\begin{quote}
\large We present a simple method for analyzing the impact of precision electroweak data above and
below the $Z$-peak on  flavour-conserving heavy new physics. 
We find that experiments have probed about ten combinations of new physics effects, 
which to a good approximation can be condensed into the effective oblique parameters
 $\hat{S},\hat{T},\hat{U},V,X,W,Y$
(we prove positivity constraints  $W,Y\ge 0$)
and three combinations of quark couplings (including a distinct parameter for the bottom).
We apply our method to generic extra $Z'$ vectors.
\end{quote}
\end{center}

\normalsize

%\end{titlepage}
\thispagestyle{empty}

%\renewcommand{\thefootnote}{(\arabic{footnote})}

%%%%%%%%%%%%%%%%%%%%%%%%%%%%%%%%%%%%%%%%%%%%%%%%%%%%%%
%%%%%%%%%%%%%%%%%%%%%%%%%%%%%%%%%%%%%%%%%%%%%%%%%%%%%%

\newpage

\section{Introduction}
\label{sec:intro}
\setcounter{equation}{0}
\setcounter{footnote}{0}

The successes of the Standard Model (SM) became so boring that various physicists
wonder if they contain an important message:
the lack of evidence for new physics pushes many proposed solutions
of the higgs mass hierarchy problem into more-or-less unnatural
corners of their parameter space.

Global fits do not provide much intuition into the origin of the strongest constraints, or
even on the number of new-physics parameters that are strongly constrained.
Here we present an efficient and simple general analysis of 
electroweak precision data using an
effective-theory description. Assuming that  new physics is somewhat
above the weak scale, its low-energy effects can be described by an
effective Lagrangian that contains leading non-renormalizable terms.
Even assuming that the new physics is generation independent (i.e.\
no new flavour physics), previous analyses identified an irreducible
set of 10 gauge-invariant operators~\cite{BS} contributing to precision
measurements at and below the $Z$-pole. This list of operators has
grown to about 20~\cite{skiba}, after that the relevance of LEP2 precision
measurements above the $Z$-pole was pointed out~\cite{STWY}.

We here show that experiments have so far precisely probed only
about 10 combinations of the 20 operators. However, if one 
follows the traditional route of  constraining new physics one must
compute all operators and then perform a global fit to
all 20 parameters: otherwise one cannot know if the new physics
corresponds to a strongly or weakly constrained combination of
higher dimensional operators.

The main aim of this paper is to develop a simpler strategy: we
identify a minimal set of parameters that are strongly constrained,
extending the $Z$-pole parameters of~\cite{PT}. In this way,
cancellations between the various operators, like the ones pointed
out in~\cite{GST}, are already built-in to this formalism. 
The data
requires almost all of these parameters to be compatible with the SM at the
{\em per mille} level.
Moreover, we want our minimal set to catch the
main features of the measurements: a reasonably accurate bound on
the scale of new physics can be extracted by just considering our
minimal set of parameters and without the necessity of a complete
analysis.

We start by identifying the sub-set of most precise measurements, mostly
performed at
$e^+e^-$ colliders (LEP1, LEP2, SLD).
Those experiments studied all $f\bar{f}$
final states, but could measure leptonic final states more
precisely than hadronic final states. We will show that the
corrections to all leptonic data can be converted into oblique
corrections to the vector boson propagators, and condensed into the
seven parameters $\hat{S}$, $\hat{T}$, $W$, $Y$, $X$, $\hat U$ and
$V$ defined in~\cite{STWY}. (Unlike in~\cite{STWY} we do not
restrict our attention to oblique new physics). Indeed, starting
with a generic set of higher-dimensional operators, one can use the
three equations of motion for $W^+,Z,\gamma$ to eliminate the three
currents involving charged leptons from the higher dimensional
operators:

\beq\label{eq:currents}
\bar{e}_L \gamma_\mu e_L,\qquad \bar{e}_R \gamma_\mu e_R,\qquad \bar{e}_L\gamma_\mu
\nu_L+\hbox{h.c.}\,.\eeq
Parameterizing the new physics in terms of
corrections to vector boson propagators is convenient because: i) in
many models the oblique parameters can be calculated
directly~\cite{arabic,STWY,LHSTWY}, without having to first
calculate the general set of induced higher dimensional operators;
ii) it is also easier to compute how the observables are affected by
oblique corrections;
iii) it allows one to unambiguously identify the most relevant corrections
to electroweak precision measurements in any generic model.

We will show that already this subset of parameters is enough to
establish the correct bound on generic models within a `typical'
20\% accuracy. Thus for most models it suffices to calculate the
seven generalized oblique parameters to establish a reasonably
accurate bound on the scale of new physics, with the caveat that the
approximation fails spectacularly if, for some reason,  new physics
is leptophobic (i.e.\ if quarks are much more strongly affected than
leptons).

A more accurate approximation is obtained by adding more parameters
in the quark sector. Basically, we keep the oblique approximation
in the ${\rm U}(1)_Y$ sector but not in the $\SU(2)_L$ sector. In
practice, this amounts to adding 2 more parameters that describe the
coupling of the left-handed quarks (which is better measured
because the larger SM coupling to the $Z$ enhances the interference
term with respect to the right-handed components). Finally, we allow
the third-generation of quarks to behave differently from lighter
quarks, and describe this possibility by adding one extra parameter:
the traditional $\varepsilon_b$~\cite{epsb}. This choice is motivated by
theoretical considerations (in many models of electroweak symmetry
breaking the top sector is special), by experimental considerations
($b$-tagging allows to probe $b$-quarks more precisely than lighter
quarks) and by phenomenological considerations (flavor universality
can be significantly violated only in the third generation).

We finally present numerical fits for our $7+2+1$ new-physics parameters,
$$\hat{S},\hat{T},\hat{U},V,X,W, Y, C_q, \delta \varepsilon_q,\delta\varepsilon_b$$
emphasizing their combinations that are most strongly constrained.
Furthermore, in section~\ref{sec:WX} we show that first principles imply
positivity constraints on $W,Y \ge 0$.

\medskip

The paper is organized as follows: in section~\ref{sec:parameters}
we introduce our formalism and identify the relevant parameters. In
section~\ref{sec:fit} we fit these parameters and compare the
results with the complete analysis, showing how accurate our
approximation typically is. In sec~\ref{sec:example} we apply the
formalism to the specific case of various extra $Z'$ bosons,
compiling present constraints. In section~\ref{sec:WX} we
demonstrate the positivity constraint on the oblique parameters $W$
and $Y$. In the appendix, we explicitly write the relation between
our parameters and a general basis of gauge-invariant operators.

%%%%%%%%%%%%%%%%%%%%%%%%%%%%%%%%%%%%%%%%%%%%%%%%%%%%%%
%%%%%%%%%%%%%%%%%%%%%%%%%%%%%%%%%%%%%%%%%%%%%%%%%%%%%%
\section{The minimal set of constrained parameters}
\label{sec:parameters}
\setcounter{equation}{0}
%\setcounter{footnote}{0}
%%%%%%%%%%%%%%%%%%%%%%%%%%%%%%%%%%%%%%%%%%%%%%%%%%%%%%

The effects of heavy new physics on precision electroweak
observables can be described by adding to the SM Lagrangian
dimension 6 operators that depend on the SM fields: the gauge bosons
$W^\pm$, $Z$ and the photon $A$, the Higgs vev $v$, the fermionic
currents $J_{ff'} = \bar f \gamma^\mu f'$, and their derivatives:
\begin{equation}
 \Lag_{\rm BSM}\, \big( W^\pm_\mu, Z_\mu, A_\mu,
\partial_\mu, v, J_{ff'} \big)\,.
\label{LBSM}
\end{equation}
We are interested here in terms that do not violate flavor and CP
(and, of course, electric charge and  color should also be
conserved). The electroweak gauge symmetry $\SU(2)_L\otimes{\rm
U}(1)_Y$, spontaneously broken by the Higgs vev, implies some
relations among the coefficients of the dimension-6 terms. There are
many such operators~\cite{BW}. After eliminating the operators that
do not affect precision data and the operators that on-shell are
equivalent to combinations of other operators, one still has to deal
with many operators: 10 if LEP2 is not included~\cite{BS}, and, including LEP2,
20 operators were considered in~\cite{skiba}.
In agreement with~\cite{GST} (where it was pointed out that two combinations
can be expressed in terms of unconstrained operators) 
we find that precision data are affected by 18 independent operators,
 listed in Appendix~\ref{app}.

\begin{figure}
$$\includegraphics[width=0.8\textwidth]{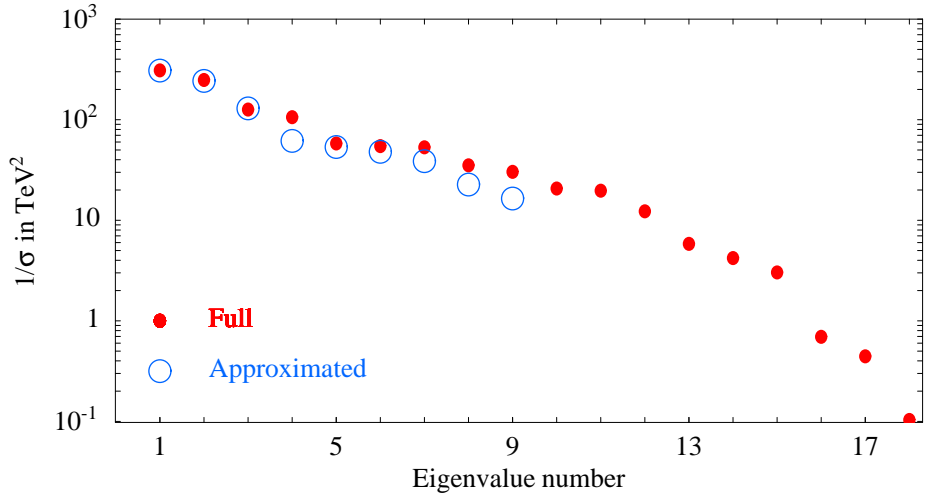}$$
\caption{\em \label{fig:sigmas} The red dots are the ordered eigenvalues of the full error matrix, that describe the sensitivity of present data (upper dots correspond to more precise combinations).
Precision data significantly constrain only about 10 new-physics effects.
The blue circles show the same eigenvalues recomputed making our simplifying approximation.}
\end{figure}

In practice, however,  many combinations of different operators are poorly constrained. 
A global analysis contains this information: one can obtain electroweak precision
bounds on a model by computing all induced higher dimensional operators. 
Our aim is to simplify this program by finding the suitable variables where
possible cancellations are manifest, and drop the unnecessary information.

In order to find the number of parameters that are strongly
constrained by the electroweak precision data we first perform the
traditional global analysis including all relevant higher dimensional
operators. In Fig.~\ref{fig:sigmas} we plot the {\em eigenvalues} of
the error matrix, computed in the uniformly-normalized basis
described in Appendix~\ref{app}.\footnote{We use the $\chi^2$ code
employed in~\cite{BS, STWY}, updating to the most recent value of
the top mass~\cite{mtop}. It agrees reasonably well with the
equivalent $\chi^2$ published in~\cite{skiba}.
We however emphasize that  the Higgs mass dependence
is not correctly approximated by keeping only the leading logarithm
analytically computed in the heavy Higgs limit, see also~\cite{STWY}.} This automatically identifies
all correlations of theoretical, experimental and accidental nature.
For example: i)  if one measurement constrains one combination of
many operators, it will appear here as one constraint; ii) if a
combination of operators does not affect any observable, it
will appear here as a zero eigenvalue. Fig.~\ref{fig:sigmas} shows
that precision data really constrain about 10
new-physics effects, and that a few constraints often dominate the global fit.
We want to find a simple
physically motivated basis for the electroweak parameters that
automatically separates the strongly constrained combinations from
the weakly constrained ones.

\medskip

We will therefore use a different approach: once a
specific set of higher dimensional operators of the form
(\ref{LBSM}) is given, we can use the equations of motion of the 3
gauge bosons $W^\pm,Z,\gamma$ to eliminate 3 fermionic currents: we
choose to eliminate the currents involving charged leptons listed in
eq.~(\ref{eq:currents}). The reason is that most of the precision
measurements have been perfomed at $e^+e^-$ colliders (LEP1, LEP2
and SLD), strongly constraining operators involving charged leptons. Neutrinos
on the other hand are experimentally more difficult to deal with than 
charged leptons. This is the reason why we have chosen to use the
equations of motion in a way that is not explicitly
$\SU(2)_L$ invariant. Muon-decay, which gives the most precise test of neutrino
couplings, is fully described by oblique couplings because it
involves charged currents and we eliminated all new physics
involving the $\bar{e}_L\gamma_\mu \nu_L$ current.

In our formalism, the most general effective Lagrangian describing new physics can be
split into two parts:
\begin{equation} \label{eq:lagrour}
\Lag_{\rm BSM} = \Lag_{\rm oblique} + \Lag_{\rm couplings} + \dots
\end{equation}
where the dots stand for terms that do not affect precision
measurements. Note again that, due to our choice for the use of the equations of
motion, $\Lag_{\rm couplings}$ will not contain any currents
involving the charged leptons. Therefore the oblique terms in
$\Lag_{\rm oblique}$ fully encode corrections to the most precisely
measured precision observables involving charged lepton final
states:
$$\alpha_{\rm em},  ~\Gamma(\mu), ~
M_Z,  ~M_W, ~ \Gamma(Z\to \ell\bar\ell),~ A^\ell_{FB}, ~A^\ell_{LR},
~ A_{\rm pol}^\tau, ~\sigma_{\rm LEP2}(e\bar e\to \ell\bar\ell), ~ee\to
ee.$$ $\Lag_{\rm couplings}$, on the  other hand, contains
corrections to the couplings of quark and neutrino currents: it
affects observables involving neutrinos and quarks~\footnote{We do
not include in the fit precision measurements of $\sigma(\nu\, {\rm
Fe})$ because they are  limited by the unprecisely known nucleon
structure: e.g.\ a strange momentum asymmetry or an isospin breaking
can account for the discrepancy with respect to the SM claimed
by~\cite{NuTeV}. Although at this stage $\Gamma(Z\to\nu \bar\nu)$ is
listed among the effects not fully described by the oblique
approximation, a detailed analysis will show that it actually is. }:
$$\Gamma(Z\to\nu \bar\nu), ~
\Gamma(Z\to q\bar q),~
A^b_{FB}, ~ A^b_{LR},~ A^c_{LR},~A^c_{FB},
\sigma_{\rm LEP2}(e\bar{e}\to q\bar{q}),~ Q_W\,.
%,~\sigma(\nu {\rm Fe})
$$
This formalism therefore allows one to clearly distinguish which
parameters are more constrained than others. This approach has
already been used in the case of models with universal new
physics~\cite{STWY} (e.g.\ gauge bosons in extra dimensions, most
little Higgs models~\cite{littlehiggs}, Higgsless models~\cite{hless}), where all corrections
involving fermions only appear in combinations proportional
to SM gauge currents. As a consequence, all fermion operators can be
completely transformed into oblique operators by using the equations
of motion for vectors. More importantly, in various concrete models
one can bypass the step of identifying the set of induced dimension
six operators: by integrating out the combinations of new-physics
vectors not coupled to fermions (rather than the heavy mass
eigenstates) directly gives the Lagrangian in terms of the oblique
parameters. Thus this method simplifies both the intermediate
computations and the final result. Here we show that this formalism
is also useful in the case of generic non-universal models (e.g.
fermions that live in different places in extra dimensions, some
Little Higgs models~\cite{simplest}, models with extra $Z'$ bosons).

In the next part of this section, we review the standard
parametrization of oblique new physics. We later present the generic
form for $\Lag_{\rm couplings}$, emphasizing the (weak) restrictions
imposed by $\SU(2)_L$-invariance, and discuss to which extent
$\Lag_{\rm couplings}$ can be neglected. In Appendix~\ref{app} we
also explicitly show how the equations of motion allow us to relate
the standard basis of $\SU(2)_L$-invariant dimension 6 operators to
our parametrization. These operators are assumed to have generic
coefficients, such that Appendix~\ref{app} applies to generic new
physics. More importantly, in section~\ref{sec:example} we show, in
a specific example of new physics (a heavy $Z'$), how one can
directly compute the full set of oblique parameters without having
to pass trough the standard basis.

\subsection{The oblique parameters}

Here we review how generic heavy new physics can affect the kinetic
terms of vector bosons, $\Pi_{33}(p^2)$, $\Pi_{30}(p^2)$,
$\Pi_{30}(p^2)$, $\Pi_{WW}(p^2)$, defined by the effective
Lagrangian
\begin{equation}
\Lag_{\rm oblique} = -\frac{1}{2} W^3_\mu \Pi_{33} (p^2) W^{3\mu}
-\frac{1}{2} B_\mu \Pi_{00} (p^2) B^\mu - W^3_\mu \Pi_{30} (p^2) B^\mu
- W_\mu^+ \Pi_{WW} (p^2) W^\mu_-\,.
\end{equation}
Since new physics is assumed to be heavy, we can
 expand the $\Pi$'s in powers of $p^2$:
\begin{equation}
\Pi (p^2) = \Pi(0) + p^2\, \Pi'(0) + \frac{(p^2)^2}{2}\, \Pi''(0) + \dots
\end{equation}
neglecting higher order terms,  that for dimensional reasons
correspond to operators of dimension higher than 6. This expansion
contains 12 parameters: 3 can be reabsorbed in the definitions of
the SM parameters $g$, $g'$ and $v$ and 2 vanish because of
electro-magnetic gauge invariance: the photon is massless and
couples to $Q= T_3+Y$. New physics is described by 7 dimensionless
oblique parameters, defined as (contrary to~\cite{STWY} we use
canonically normalized kinetic terms)
\begin{equation}
\begin{array}{c}\displaystyle
\hat S = \frac{g}{g'} \Pi_{30}'\,, \quad \hat T = \frac{\Pi_{33} - \Pi_{WW}}{M_W^2}\,, \quad W = \frac{M_W^2}{2} \; \Pi''_{33}\,, \quad Y = \frac{M_W^2}{2} \; \Pi''_{00}\,,\\ \displaystyle
\hat U =\Pi'_{WW} -  \Pi'_{33} \,, \qquad V = \frac{M_W^2}{2} \;
(\Pi''_{33} - \Pi''_{WW})\,, \qquad X = \frac{M_W^2}{2} \;
\Pi''_{30}\,,
\end{array}
\end{equation}
where all $\Pi$'s are computed at $p^2=0$. These parameters correct
the propagators of the gauge bosons, affecting the precision
observables. Only 6 combinations actually enter observables
involving charged leptons: in particular, only the combination $\hat
U - V$. $Z$-pole precision data can be encoded in the
$\varepsilon$'s of~\cite{eps}. Low energy data do not depend on
$\hat{U}, V$. The $e\bar{e}\to f\bar{f}$ cross sections measured at
LEP2  are dominantly affected by  $Y$, $W$ and $X$~\cite{STWY}.

Using $\SU(2)_L$-invariance one can show that $V\ll \hat{U}\ll
\hat{T}$ and $X\ll \hat{S}$: in the case of universal new physics,
the sub-leading form factors  $\hat{U},V,X$ can therefore be
neglected and new physics is fully described by $\hat{S},\hat{T},W,
Y$~\cite{STWY}. This argument however does not apply in our case,
where the same parameters are applied in  a different context: to
describe how generic heavy new physics (not necessarily universal)
affects observables that only involve charged leptons and vectors. To reach the
basis in which charged-leptonic data are condensed into vector
propagators we made a transformation which is not
$\SU(2)_L$-invariant. As a consequence all oblique parameters
generically arise at leading order.

%%%%%%%%%%%%%%%%%%%%%%%%%%%%%%%%%%%%%%%%%%%%%%%%%%%%
\subsection{Vertex corrections} \label{sec:vertex}
%%%%%%%%%%%%%%%%%%%%%%%%%%%%%%%%%%%%%%%%%%%%%%%%%%%

Here we present the effective Lagrangian that describes new-physics
corrections to $Z,\gamma$ couplings, taking into account a) that we
eliminated currents involving charged leptons; b) that new physics
is heavy, allowing a low-energy expansion in momenta; c)
electromagnetic gauge invariance. A convenient parametrization is:
\begin{equation}\label{eq:vert}
\Lag_{\rm couplings} = \sum_f (\bar f \gamma^\mu f) \left[ e\, A_\mu\, \frac{C^\gamma_f}{M_W^2}\, p^2
+ \sqrt{g^2+{g'}^2}\, Z_\mu \left( \frac{C^Z_f}{M_W^2}\, (p^2 -
M_Z^2) + \delta g_{f} \right) \right]\,, 
\end{equation}
%Here the currents are defined as $J_{f,\mu} =\bar{f} \gamma_\mu f$ for
where $f={u_L,d_L, u_R,d_R,\nu_L}$, and higher orders in the
momentum again correspond to subleading effects due to operators
with dimensions greater than 6. The $\delta g$'s are corrections to
on-shell $Z$ couplings, tested by measurements at the $Z$-pole. The
$C^\gamma$ and $C^Z$ are equivalent to 4-fermion contributions to
$e^+ e^- \rightarrow q \bar q$: the $p$-dependence cancels the
propagator of the gauge boson, and we are left with a constant
($p$-independent) contribution. They affect LEP2, atomic parity
violation, etc. For the neutrinos only $\delta g_{\nu_L}$ is
measured via the invisible decay ratio of the $Z$.

$\SU(2)_L$ invariance implies some mild restrictions on these vertex parameters:
\begin{enumerate}
\item  As shown in Appendix~\ref{app}, $\delta g_{L\nu}$ is fixed in terms of oblique parameters as:
\begin{equation}
\delta g_{L \nu}= V - \frac{1}{2} \hat{U} - \tan \theta_W X\,.
\end{equation}
Notice that it depends on a different combination of $\hat U$ and
$V$ than the one  entering corrections to the gauge boson
propagators. This means that considering all the 7 oblique
parameters defined in the previous subsection is enough to include
the relevant neutrino measurements.

\item
In the quark sector we apparently have 12 new parameters: $\delta
g_{L,R \, u,d}$ and $C^{Z,\gamma}_{L,R \, u,d}$. However only 11 of
them are independent, and correspond to the 11 quark operators
of~\cite{skiba}. Indeed, as explicitly shown in Appendix~\ref{app},
the following relation holds between the 4-fermion coefficients of
the left-handed quarks:
\begin{equation}
(C^\gamma_{dL}-C^\gamma_{uL}) = \cos^2 \theta_W (C^Z_{d L}-C^Z_{u L}) + \frac{X}{\tan \theta_W}.
\label{relations}
\end{equation}

\end{enumerate}

%%%%%%%%%%%%%%%%%%%%%%%%%%%%%%%%%%%%%%%%%%%%%
\subsection{A simple approximation}
%%%%%%%%%%%%%%%%%%%%%%%%%%%%%%%%%%%%%%%%%%%

We can now proceed with the final counting.
We have 18 independent coefficients: the 7 oblique parameters
$\hat{S},\hat{T},\hat{U}, V,X,Y,W$, 4 $\delta g$'s for the
quarks, 4 quark $C^Z_q$'s and 3 independent $C^\gamma_q$'s.
The counting agrees with the results of~\cite{GST},
that shows how 2 combinations of the 21 operators of~\cite{skiba}
can be eliminated. (18 arises as $21-2-1$:
one further operator, that only affects $e^- e^+\to W^+ W^-$,
is ignored here because we do not view this as a `precision' measurement.
This view is corroborated by the numerical results of~\cite{skiba,GST}.)
In other words, the unconstrained combinations pointed out in~\cite{GST}
are automatically eliminated in our formalism.

Our basis makes a clear separation of which parameters contribute to
which measurements. Corrections to  observables involving leptons
only  are expressed in terms of the seven oblique parameters (which
as we have seen also include neutrinos). Observables involving
quarks in the final state at the $Z$-pole  involve in addition only
the four $\delta g$'s. The $C^{\gamma,Z}_q$'s are only necessary for
$\sigma(e\bar{e}\to q\bar{q})$ at LEP2 and atomic parity violation.

As leptonic final states are generically better measured than
hadronic ones, this separation already suggests that describing the
precision measurements in terms of only the 7 oblique parameters
could be a reasonable approximation (oblique approximation). In the
next section we will check numerically that this indeed happens.
This approximation also includes the constraints on neutrinos.

In order to be more accurate, we want to add a minimal set of parameters
describing corrections in the hadronic sector. In fact, not all the
quark observables are well measured, so that only a small subset of
parameters  will actually contribute most strongly to the bound. At
the $Z$-pole, the better measured quantity is the hadronic branching
ratio of the $Z$. It depends on the combination:
$$
g^{\rm SM}_{qL}\, \delta g_{qL} + g^{\rm SM}_{qR}\, \delta g_{qR}\,.
$$
Due to the fact that the couplings of the right-handed components to
the $Z$ are generically smaller than the couplings of the
left-handed component (by a factor of 0.18 for the down type quarks,
and 0.44 for the up type), we expect in general that only the
corrections involving left-handed quarks will be relevant. Moreover,
when the contribution of up and down quarks are summed, the result
is proportional to:
$$
\delta g_{uL} - \delta g_{dL} -  \frac{\tan^2 \theta_W}{3}\, (\delta g_{uL} +
\delta g_{dL})\,,
$$
so that the difference between the two parameters seems to be more
relevant than the sum.

Similar arguments apply for the hadronic cross section measured at
LEP2. The main difference is the presence of interference with the
SM diagram with a photon exchange,  and the presence of 4-fermion
operators. We first notice that the interference with the photon is
generically suppressed by the gauge coupling $e$ versus $\sqrt{g^2 +
{g'}^2}$: this results in a suppression of order $\sin \theta_W$. The
contribution of the $\delta g$'s will therefore enter in the same
way as in the hadronic branching ratio. A very similar argument can
be applied to the 4-fermion contribution, so that only the
combinations $C^Z_{Lu} - C^Z_{Ld}$ and $C^\gamma_{Lu} -
C^\gamma_{Ld}$ are constrained: as already mentioned in
(\ref{relations}) these two parameters are related to each other, so
that they correspond to a single parameter. From this rough argument
we can thus infer that 2 parameters will be most relevant in the
quark sector:
\begin{eqnsystem}{sys:sss}
\delta \varepsilon_q & = & \delta g_{uL} - \delta g_{dL}\,, \\
\delta C_q  & = & C^Z_{uL} - C^Z_{dL}\,.
\end{eqnsystem}
Again, in the next section we will numerically show that this is indeed the case.

Until now we have assumed flavor universality including the third
generation, and in particular the bottom quark. However, in many
models of electroweak symmetry breaking the third generation of
quarks is special due to the heavyness of the top quark, and it is
differently affected by new physics. For this reason, we will relax
the flavor universality for the bottom quark, and deal with it
separately. This is also necessary since the bottom final state is
well measured.
At LEP1, only $\delta g_{bL}$ is
well measured, because the SM coupling of the right-handed component
is smaller, thus we can define:

\beq \delta g_{bL} = -\frac{1}{2} \delta \varepsilon_b\,; \eeq
here the parameter $\delta \varepsilon_b$ coincides with the standard
definition given in~\cite{epsb}. Notice that the anomalous $A_{FB}^b$
measurement gives a subleading contribution to the determination of
$\delta\varepsilon_b$.
The cross section $\sigma (e\bar e \to b\bar
b)$ at LEP2 also depends on a combination of 4-fermion operators.
In general an extra parameter should also be added to the fit: however, in model of electroweak symmetry breaking involving the top quark, we expect corrections to $\delta \varepsilon_b$ to be more important.
The reason is that the 4-fermion operators with the bottom will also involve couplings of new physics with the electron, already tightly constrained by the oblique parameters.
This is the case, for example, in models with dynamical symmetry breaking~\cite{contino}, gauge-Higgs unification~\cite{ghu} or Higgsless models~\cite{hless}.
Thus, in order to simplify the analysis, we will approximate a flavour-universal contribution to the bottom 4-fermion operators.
In this way, only one parameter is sufficient to describe the bottom.

%%%%%%%%%%%%%%%%%%%%%%%%%%%%%%%%%%%%%%%%%%%%
\section{Global fit}
\label{sec:fit}
\setcounter{equation}{0}
%\setcounter{footnote}{0}
%%%%%%%%%%%%%%%%%%%%%%%%%%%%%%%%%%%%%%

In this section we study the fit of the precision electroweak
measurements, and show that the approximations proposed in the
previous section are actually sensible, and give a sufficiently
reliable bound on generic models of new physics.
One can express all the
observables in terms of the following 18 parameters:
the 7 oblique parameters $\hat S$, $\hat T$,
$\hat U$, $W$, $Y$, $V$ and $X$, 4 corrections to the couplings of
the $Z$ with quarks $\delta g_{uR}$, $\delta g_{dR}$, $\delta
g_{uL}$, $\delta g_{dL}$, and 7 4-fermion parameters
(4 involving
right-handed quarks $C^\gamma_{uR}$, $C^\gamma_{dR}$, $C^Z_{uR}$,
$C^Z_{dR}$, and 3 involving left-handed quarks $C^Z_{uL}$,
$C^Z_{dL}$, and $C^\gamma_{uL} + C^\gamma_{dL}$). Note that in doing
this we are not yet introducing any approximation: we are just
choosing a  particular basis for the dimension 6 operators affecting
electroweak precision observables.

The two approximations  we want to pursue are the following:  first
we consider only the 7 oblique parameters $\hat S$, $\hat T$, $\hat
U$, $W$, $Y$, $V$ and $X$ (oblique approximation) and set all the
others to zero: this allows us to exactly describe the observables
only  involving vectors and leptons (charged and neutrinos),
but, in general, does not correctly describe  corrections to quark observables.
Next, as argued in the previous section, in the quark sector two parameters should
have the strongest effect on the bound on new physics. They are
related to corrections to the couplings to the $Z$ and 4-fermion
operators involving left-handed components, $\delta \varepsilon_q$
and $\delta C_q$.

\medskip

\begin{figure}[tb]
\begin{center}$$\hspace{-0.06\textwidth}
\includegraphics[width=0.4\textwidth]{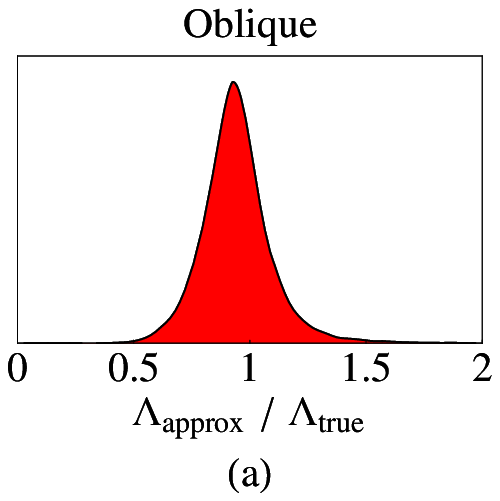}
\hspace{-1cm}
\includegraphics[width=0.4\textwidth]{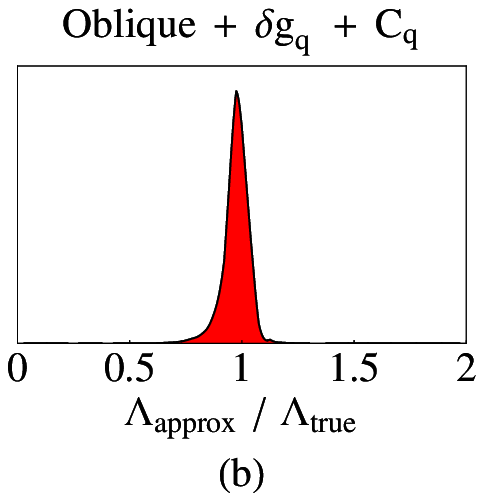}
\hspace{-1cm}
\includegraphics[width=0.4\textwidth]{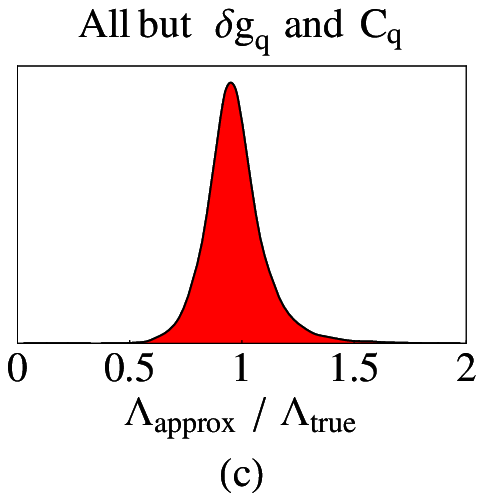}$$
\caption{\em Distibutions of the ratios between the approximate over
true  bound in various approximations. In the first {\it ``oblique''} panel
we include in the fit only $\hat S$, $\hat T$,
$\hat U$, $W$, $Y$, $V$ and $X$. In the second panel we add the two parameters
$\delta C_q$ and $\delta \varepsilon_q$ for the quarks. Finally we include all the parameters except $\delta C_q$ and $\delta \varepsilon_q$.}
\label{fig:histo}
\end{center}
\end{figure}

We now check how good our approximations are for guessing
the bound on the scale $\Lambda$ of new physics in generic models. To do
that, we generated many random models
by writing each parameter as $r/\Lambda^2$, where
$-1\le r \le 1$ are random numbers.
This is an reasonably arbitrary procedure.
We then extract the bound on $\Lambda$ both from the exact fit and the approximate
fits. The result is graphically shown in Fig.~\ref{fig:histo}: in
case (a) we show the  oblique approximation; in (b) we add the
two parameters $\delta C_q$ and $\delta \varepsilon_q$ for the
quarks to the oblique parameters; in (c) we include all the
parameters except $\delta C_q$ and $\delta \varepsilon_q$. In the
following table we report, for the same cases, the average value and the variance of
$\Lambda_{\rm approx}/\Lambda_{\rm true}$.

\vspace{0.3cm} \begin{center}
\begin{tabular}{c|c}
Approximation &  $\Lambda_{\rm approx}/\Lambda_{\rm true}$ \\
\hline
Oblique & $0.95 \pm 0.16$ \\
Oblique plus $C_q$, $\delta \varepsilon_q$   & $0.98 \pm 0.06$ \\
All but $C_q$, $\delta \varepsilon_q$  & $0.98 \pm 0.15$
\end{tabular}
\end{center} \vspace{0.3cm}
We see that the oblique approximation is already
reasonable:  in most of the cases the approximate bound is less than
$25 \%$ away from the correct one. Adding the two parameters $\delta
C_q, \delta \varepsilon_q$ improves the approximation significantly:
in more than $90 \%$ of the cases the approximate bound reproduces
the exact one within $10 \%$. Furthermore, it is important to notice
that considering a fit where all the parameters except $\delta C_q ,
\delta \varepsilon_q$ are added does not improve much the
approximation with respect to the oblique case. This is telling
us that in the quark sector it is indeed $\delta C_q$ and $ \delta
\varepsilon_q$ which are the most constrained parameters, while all
the others are much less constrained (and mostly negligible for
establishing a reliable bound on the scale of new physics). The
arguments we have discussed in section \ref{sec:vertex} thus find a
quantitative verification here. Out of the 18 initial parameters
only 9 are truly constrained. The remaining 9 can be safely
neglected.

Fig.~\ref{fig:sigmas} compares the eigenvalues of the full error
matrix with the eigenvalues recomputed using our simplified approximation
(using of course the same normalization in the two cases).
We see that the approximation catches the main constraints,
ignoring the remaining weakly constrained combinations.
We do not show the full eigenvalues extracted from the global fit of~\cite{skiba},
that show a similar level of agreement.

\bigskip

We now present how data determine our 10 parameters by
presenting  the `eigenvectors' of the global  $\chi^2$,
i.e.\ we show the orthogonal combinations 
that have been determined
with no statistical correlation with the other combinations, such that
a model is excluded if any one of these combinations contradicts experimental data.
We order them starting from the most precise ones.
They are:
\beq R \cdot \left(\begin{array}{c} \hat{S}\\ \hat{T} \\ \hat{U} \\ V \\ W \\ X \\ Y  \\ \delta C_q \\ \delta \varepsilon_b \\ \delta\varepsilon_q
\end{array}\right)= 10^{-3}
\left(\begin{array}{c}
-0.04 + 0.54 \ell \pm 0.21\\ 
+0.13 + 0.08 \ell \pm 0.43\\ 
+0.41 + 0.21 \ell \pm 0.50\\ 
+0.16 + 0.72 \ell \pm 0.54\\ 
-0.36 - 0.33 \ell \pm 0.75\\ 
0 + 0.16 \ell \pm 1.2\\ 
-0.9 - 0.12 \ell \pm 1.5\\ 
-5.6 - 0.31 \ell \pm 2.0\\ 
-0.4 + 0.18 \ell \pm 8.7\\ 
-26 + 0.66 \ell \pm 18
\end{array}\right)
\eeq
where
the factor $\ell = \ln (m_h/M_Z)$ encodes the approximate dependence on the Higgs mass
and the orthogonal matrix $R$ equals
$$R = 10^{-3} \left(
\begin{array}{cccccccccc}
 -404 & 353 & -133 & 173 & 137 & -753 & 276 & 4 & 18 & 27 \\
 -245 & -19 & 492 & -747 & 30 & -37 & 280 & 15 & -40 & -235 \\
 -16 & 208 & 146 & -152 & -724 & -224 & -407 & 319 & 33 & 260 \\
 -222 & 691 & -76 & 5 & -120 & 550 & 285 & -129 & 55 & 216 \\
 -17 & -330 & 177 & -36 & 114 & -31 & 273 & -12 & 1 & 876 \\
 3 & 232 & -7 & -283 & 303 & -118 & -589 & -581 & -175 & 209 \\
 -42 & -68 & 132 & 31 & -44 & -37 & -66 & -288 & 939 & -33 \\
 -203 & -200 & 350 & 375 & -445 & -9 & 126 & -587 & -282 & -124 \\
 -642 & -381 & -575 & -219 & -161 & 147 & -112 & -41 & 9 & 11 \\
 519 & 0 & -458 & -341 & -329 & -199 & 376 & -337 & -1 & 2
\end{array}
\right).
$$
The two last combinations have large uncertainties and can be ignored.
The flavour-universal limit is obtained by setting $\delta \varepsilon_b = \delta \varepsilon_q$.

%%%%%%%%%%%%%%%%%%%%%%%%%%%%%%%%%%%%%%%%%%%%%%%%%%%%%%
%%%%%%%%%%%%%%%%%%%%%%%%%%%%%%%%%%%%%%%%%%%%%%%%%%%%%%

\section{Example: a generic $Z'$}
\label{sec:example}
\setcounter{equation}{0}

We now apply our results to  a specific concrete example: a generic
heavy non-universal $Z'$ vector boson, with mass $M_{Z'}$, gauge
coupling $g_{Z'}$ and gauge charges $Z_X$ under the various SM
fields $X=\{ H,E,L,Q,U,D\}$.
The
parameters
defined in section \ref{sec:parameters}  can be computed in various ways.
One can integrate out the heavy mass
eigenstate, obtaining a set of effective operators that can be converted into
our parameters using the expressions in Appendix \ref{app}.
A  simpler technique~\cite{arabic} allows to directly compute our parameters.
In the specific case of a $Z'$ this technique was described in  section~7 of \cite{LHSTWY}:
it consists of integrating out the combination of $Z'$ and $Z$
(which, in general, is not a mass eigenstate)
that does not couple to charged leptons.
Operatively, one rewrites the Lagrangian in terms of
\begin{equation}
  \tilde B_\mu = B_\mu - \frac{g_{Z'}Z_E}{g' Y_E} Z'_\mu, \;\;\;\;\;\;\; \tilde W_\mu^3 = W_\mu^3
   - \frac{2 g_{Z'}}{g Y_E} (Z_E Y_L-Z_L Y_E) Z'_\mu
\end{equation}
such that in the new basis $Z'$ no longer couples to charged leptons and can be
integrated out without generating any operator involving charged leptons. 
One can then directly  extract our 9 parameters from the effective Lagrangian, since it
already is in the form of eq.~(\ref{eq:lagrour}).
The explicit result is:
\begin{eqnsystem}{sys:Z'}
\hat S &=& \frac{2 M_W^2 g_{Z'}^2}{g^2 g'^2 M_{Z'}^2} (Z_E-Z_H+Z_L)(g^2 Z_E +g'^2 (Z_E+2Z_L)) \,,\\
\hat T &=& \frac{4 M_W^2 g_{Z'}^2}{g^2 M_{Z'}^2} (Z_E-Z_H+Z_L)^2 \,,\\
\hat U & = & \frac{4 M_W^2 g_{Z'}^2}{g^2 M_{Z'}^2} (Z_E-Z_H+Z_L)(Z_E+2Z_L) \,,\\
W &=& \frac{M_W^2 g_{Z'}^2}{g^2 M_{Z'}^2} (Z_E+2Z_L)^2 \,,\\
Y &=&  \frac{M_W^2 g_{Z'}^2}{g'^2 M_{Z'}^2} Z_E^2 \,,\\
V &=&  \frac{M_W^2 g_{Z'}^2}{g^2 M_{Z'}^2} (Z_E+2Z_L)^2 \,,\\
X &=& - \frac{M_W^2 g_{Z'}^2}{g g' M_{Z'}^2} Z_E (Z_E+2Z_L) \,, \\
\delta \varepsilon_q &=& \frac{2 M_W^2 g_{Z'}^2}{g^2 M_{Z'}^2} Z_H (Z_E+2 Z_L)\,, \\
\delta C_q &=& \frac{2 M_W^2 g_{Z'}^2}{(g^2+g'^2) M_{Z'}^2} (Z_E+2Z_L) (Z_E+Z_L) \,.
\end{eqnsystem}
It is important to notice a point missed  in section~7 of \cite{LHSTWY}:
 $\hat U$, $V$ and $X$ are not subdominant with respect to $\hat{S}$, $\hat{T}$, $W$, $Y$.
(The bounds presented here numerically differ from the ones in  \cite{LHSTWY} also because
we here updated the measurement of the top mass~\cite{mtop}).
One can check that only with the correct full expressions of eq.~(\ref{sys:Z'})
the corrections to the parameters $\delta \varepsilon_{1,2,3}$
that summarize LEP1 observables are all proportional to $Z_H$ and
therefore all vanish if the Higgs is neutral under the heavy $Z'$.
This must happen because  $Z_H=0$ means no $Z/Z'$ mixing
and the $Z'$ manifests itself only  as 4-fermion operators invisible at LEP1
and dominantly constrained by LEP2.

\begin{table}[t]
$$
  \begin{array}{cc|cccccc|ccc}
\hbox{U(1)}& \hbox{universal?}& Z_H & Z_L & Z_D & Z_U & Z_Q & Z_E & \hbox{full} &\hbox{approx} & \hbox{oblique} \\ \hline
H & \hbox{yes} & 1 & 0 & 0 & 0 & 0 & 0 & 6.7 & 6.7 & 6.7\\ 
B' & \hbox{yes} & \frac{1}{2} & -\frac{1}{2} & \frac{1}{3} & -\frac{2}{3} & 
\frac{1}{6} & 1 & 6.7 & 6.7 & 6.7\\ 
B'_F & \hbox{yes} & 0 & -\frac{1}{2} & \frac{1}{3} & -\frac{2}{3} & 
\frac{1}{6} & 1 & 4.8 & 4.8 & 4.8\\ 
B-L & \hbox{no} & 0 & -1 & -\frac{1}{3} & -\frac{1}{3} & \frac{1}{3} & 1 & 
6.7 & 7.1 & 7.1\\ 
L & \hbox{no} & 0 & 1 & 0 & 0 & 0 & -1 & 6.3 & 7.1 & 7.1\\ 
10 & \hbox{no} & 0 & 0 & 0 & 1 & 1 & 1 & 2.5 & 2.9 & 3.4\\ 
5 & \hbox{no} & 0 & 1 & 1 & 0 & 0 & 0 & 3.8 & 3.2 & 5.6\\ 
Y & \hbox{no} & \frac{2}{3} & 1 & 1 & -\frac{1}{3} & -\frac{1}{3} & 
-\frac{1}{3} & 4.8 & 5.0 & 6.0\\ 
16 & \hbox{no} & 0 & 1 & 1 & 1 & 1 & 1 & 4.4 & 4.7 & 6.5\\ 
\hbox{SLH} & \hbox{no} &  \multicolumn{6}{c|}{\hbox{Simplest little Higgs~\cite{simplest}}} & 2.7 & 
2.5 & 2.7\\ 
\hbox{SU6} & \hbox{no} &  \multicolumn{6}{c|}{\hbox{Super little Higgs~\cite{superlittle}}} & 3.1 & 3.3 
& 3.3\\  \end{array}$$
  \caption{\em $99\%$\CL\ bounds on the ratio $M_{Z'}/g_{Z'}$ in {\rm TeV} for a set of frequently studied $Z'$.}
  \label{tab:famZprimes}
\end{table}

\begin{figure}[tb]
\begin{center}$$\hspace{-0.03\textwidth}
\includegraphics[width=1.05\textwidth]{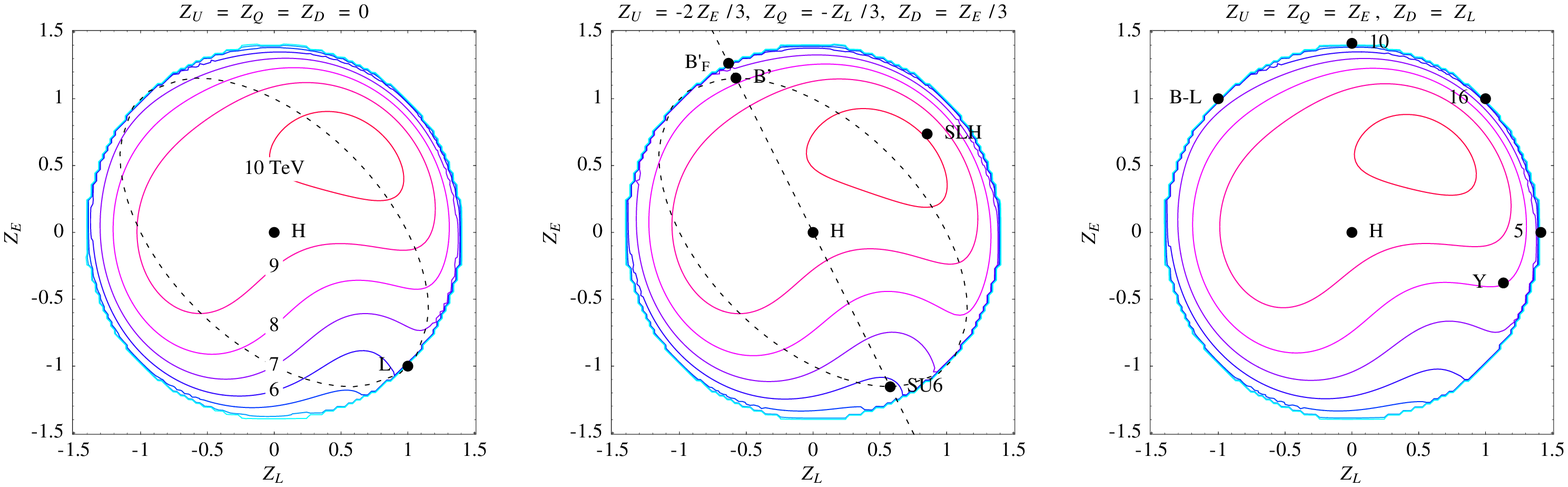}$$
\caption{\em Bounds on $M_{Z'}/g_{Z'}$ in $\TeV$ at $99\%$\CL\ for
different $Z'$ models. Their effect dominantly depends on the charge
of the Higgs and the leptons: we here assume the normalization
$Z_L^2 + Z_E^2 + Z_H^2 = 2$ such that $Z_H=0$ at the boundary of the
circles. The three plots, done assuming different sets of quark
charges (zero, universal-like and SU(5)-unified) are almost
identical, confirming the validity of an approximate analysis. The
dashed line corresponds to a universal $Z'$, and the dashed ellipse
to a $Z'$ compatible with SM Yukawa couplings. The dots show some
well-known $Z'$s.} \label{fig:Z'}
\end{center}
\end{figure}

In table \ref{tab:famZprimes} we report the $99\%$\CL\ bounds on  $M_{Z'}/g_{Z'}$ for a set of  $Z'$s,
theoretically motivated by extra dimensions, unification models, little Higgs models.\footnote{We
presented the results hiding a  technical problem. 
We performed two different global fits: in the operator basis, and
in the oblique basis. The simpler  oblique analysis naturally
allows to include minor effects.
The minor difference between the two $\chi^2$ is comparable to
the accuracy of our approximation, such that in the table we
compensated for this.}
We compare the bound obtained by performing an exact fit, that
includes the effects of all the 18 relevant parameters, an
approximate fit including the 9 parameters, and the purely oblique
approximation. It is interesting to notice that the approximate
bounds reproduce the exact one accurately in almost all the cases.
There are few exceptions where the effect of quarks is relevant, and
the oblique bound is overestimated. On the other hand, the
9-parameter approximation is always successful.

Fig.~\ref{fig:Z'} shows iso-contours of bounds on $M_{Z'}/g_{Z'}$ (computed assuming a light Higgs)
that approximately apply to
all $Z'$.  Indeed the constraint dominantly depends only on the leptonic and Higgs $Z'$ charges:
$Z_H,Z_L,Z_E$. We here fixed their arbitrary overall normalization by assuming
$Z_H^2 + Z_L^2 +Z_E^2=2$. Without loss of generality we can choose $Z_H\ge 0$, such
that all the information lies on the surface of a half-sphere, and is plotted in fig.~\ref{fig:Z'}.
The different panels show three different arbitrary choices for the quark $Z'$ charges:
vanishing (left panel), universal (middle panel), SU(5)-unified (right panel).
Each panel shows the exact bound on $M_{Z'}/g_{Z'}$:
one sees that there are very minor differences between the bounds in the three panels,
confirming that leptonic data dominate the present global fit.
The dots show the locations of the theoretically-motivated $Z'$ listed in table~\ref{tab:famZprimes}.
The dashed lines show special sub-classes of $Z'$: universal $Z'$ (oblique line)
and $Z'$s that do not forbid the SM Yukawa couplings (ellipse).
For example, only two $Z'$ have both these properties:
a) the one denoted as $B'$: a duplicate of the SM hypercharge;
b) the one denoted as  `SU6' $Z'$, that arises in little-Higgs models~\cite{LHSTWY}.

\section{Proof for $W,Y\ge 0$}\label{sec:WX}
\setcounter{equation}{0}

So far, the oblique parameters $W,Y$ have been computed in various
models (in extra dimensions, Higgsless models, and litte Higgs at
tree level~\cite{STWY, LHSTWY}, and in supersymmetry~\cite{SUSYSTWY}
and Minimal Dark Matter~\cite{MDM} at one-loop level). In all of
these cases it has been found that $W,Y\ge 0$. Next we discuss the
general reason behind this result. The K\"allen--Lehmann
representation implied by unitarity~\cite{book} tells us that propagators can be
written as \beq \frac{1}{\Pi(p^2)} =\int_0^\infty d m^2
\frac{\rho(m^2)}{p^2 - m^2 - i\varepsilon}\qquad\hbox{with} \qquad
\rho(m^2)\ge 0.\eeq One can compute $\Pi''(0)$ and write in an
appropriate form such that positivity is manifest:
$$ \Pi''(0)=
\frac{\int\!\int dm_1^2 dm_2^2\,\rho(m_1^2)\rho(m_2^2)
(m_1^2-m_2^2)^2/m_1^6 m_2^6}{[\int dm^2 \,\rho(m^2)/m^2]^3}
\ge 0.$$

We could similarly prove that $\Pi'(0)\ge 0$, and this indicates a
potential caveat. The K\"allen--Lehmann representations applies to
correlators of gauge invariant operators. In models where  the SM
gauge group is a subgroup of some larger non-abelian gauge group the
relevant propagators are not gauge-invariant quantities: they can
have matrix elements with unphysical negative-norm states, possibly
giving $\Pi''(0)<0$. As well known, this is indeed what happens in
the case of $\Pi'(0)$, that contributes to the $\beta$-function of
gauge couplings: non-abelian vectors negatively contribute to the
$\beta$-function. Littlest-Higgs models with $T$-parity~\cite{Tparity} might
realize this caveat: the one-loop corrections to physical
observables must be computed including the full gauge-invariant set
of oblique, vertex and box diagrams.

\section{Conclusions}

We presented a simple and efficient general analysis of the
constraints on heavy new physics from electroweak precision data measured below, at
and above the $Z$-peak. 
We found that, out of a complete basis of 18 independent operators,
precision data significantly constrain only about 10 combinations of
new-physics parameters, see fig.~\ref{fig:sigmas}.
We have condensed the dominant precision data into 7
generalized oblique parameters $\hat{S},\hat{T},\hat{U},V,X,Y,W$
(that fully describe how new physics affects vectors and leptons),
plus two parameters that describe the main corrections involving quarks:
$\delta \varepsilon_q$, that describes corrections to the on-shell $q\bar{q}Z$ vertex,
and $C_q$, that describes the size of $e\bar{e}q\bar{q}$ four fermion operators. 
A 10th parameter, the traditional $\delta \varepsilon_b$, is necessary if (as in
most models) third generation quarks have unique properties. 

We have shown that in most cases the simple oblique
approximation (where only the seven oblique parameters are turned
on) reasonably estimates  the constraints on new physics,
and that adding all 9 (or 10) parameters gives a bound that typically is within
10\% of the exact bound. We have shown how to calculate these
parameters from a generic set of higher dimensional operators, and
emphasized that an added advantage of our parameters is that in many
cases they can be directly computed via integrating out proper
combinations of heavy new physics. We applied our
methods giving approximate bounds on generic $Z'$s (see fig.~\ref{fig:Z'}), 
and compared them with exact results in the specific cases of frequently-studied $Z'$
(see table~\ref{tab:famZprimes}).

Finally, we have shown that first principles demand 
positivity constraints $W,Y\ge 0$ on these oblique parameters.

\footnotesize

%%%%%%%%%%%%%%%%%%%%%%%%%%
\section*{Acknowledgments}
We thank R.\ Rattazzi for very useful discussions.
G.C. also thanks Graham Kribs for the organization of the UltraMini Workshop at the University of Oregon (Eugene), where part of this work was completed.
The research of G.C. and C.C. is supported in part by the DOE OJI grant DE-FG02-01ER41206
and in part by the NSF grants PHY-0139738  and PHY-0098631.
The work of G.M. is supported in part 
by the Department of Energy grant
DE-FG02-91ER40674.
The research of A.S.\ is supported in part by the European Programme
`The Quest For Unification', contract MRTN-CT-2004-503369.

\normalsize

\appendix
%%%%%%%%%%%%%%%%%%%%%%%%%%%%%%%%%%%%%%%%%%%%%%%%%%%%%%%%%%%%%%%%%%%%%%%
%%%%%%%%%%%%%%%%%%%%%%%%%%%%%%%%%%%%%%%%%%%%%%%%%%%%%%%%%%%%%%%%%%%%%%%
\section{The minimal parameters for a general Lagrangian}
\label{app} \setcounter{equation}{0} \setcounter{footnote}{0}

In this appendix
we explicitly show how to transform a Lagrangian with generic $\SU(2)_L$-covariant
dimension 6 operators to the parametrization advocated here.
To start, we fix the notations for the dimension 6 operators. We define the Higgs and fermion
 currents as

\begin{equation}
J_{\mu H} =  H^\dagger i \mathcal{D}_\mu H,\qquad
J_{\mu H}^a = H^\dagger  \tau^a i\mathcal{D}_\mu H,\qquad
J_{\mu F} = \sum \bar{F} \gamma_\mu F,\qquad
J_{\mu D}^a = \sum \bar{D}\gamma_\mu \tau^a D
\end{equation}
where $\tau^a$ are the Pauli matrices (normalized such that $\mbox{Tr}\, (\tau^a \tau^b) = 2 \delta^{ab}$), %with eigenvalues $\pm 1$),
$F = \{E,L,Q,U,D\}$ and $D=\{L,Q\}$ and the currents are summed over
the three flavors. In our notation the hypercharges are $Y_E=1$,
$Y_L=-1/2$, $Y_U = -2/3$, $Y_D = 1/3$, $Y_Q = 1/6$, $Y_H = 1/2$ and $\langle H\rangle = (0,v)$. As discussed
in section~\ref{sec:parameters}, we  split doublets $D=(u,d)$ in
components and define
$$J_{\mu D} = J_\mu^{u_L} + J_\mu^{d_L} = \bar{u}_L \gamma_\mu u_L + \bar{d}_L \gamma_\mu d_L
,\qquad
J^-_{\mu D} = \bar{u}_L \gamma_\mu d_L,\qquad
J^+_{\mu D} = \bar{d}_L\gamma_\mu u_L
$$
To shorten the notation, we define $s = \sin \theta_W$, $c = \cos \theta_W$ and $t=\tan \theta_W$.

We start from a complete list of dimension 6 operators that are relevant for the precision measurements at LEP:
following the notation of~\cite{BS,skiba}, the operators are
\begin{itemize}
\item 7 operators involving one fermion current (vertex operators):
\begin{equation}
\mathcal{O}_{HF} =  J_{\mu H} J_{\mu F}+ \hbox{h.c.} =
 - 2 \frac{M_W^2}{g^2} (g W^3_\mu - g' B_\mu) J_f^\mu \, + \cdots \\
\end{equation}
 where $F=\{L,E,Q,U,D\}$ and
\begin{equation}
\mathcal{O}'_{HD} =   J_{\mu H}^a J^a_{\mu D}+ \hbox{h.c.} =
2 \frac{M_W^2}{g^2} (g W^3_\mu - g' B_\mu) (J_\mu^{u_L} - J_\mu^{d_L} ) + 4 \frac{M_W^2}{g^2} \frac{g}{\sqrt{2}} \left( W^+_\mu J^-_{D\mu} + \hbox{h.c.} \right) +\cdots
\end{equation}
where $D=\{L,Q\}$.

\item 11 operators involving two fermion currents (4-fermion operators):
\begin{eqnarray}
\mathcal{O}_{FF'} &=& \frac{J_{\mu F} J_{\mu F'}}{1+\delta_{FF'}}\\
\mathcal{O}'_{DD'} &=&
 \frac{J^a_{\mu D} J^a_{\mu D'}}{1+\delta_{DD'}}=
  \frac{(J^{u_L} - J^{d_L})\cdot (J^{u'_L}-J^{d'_L})+
  2 (J^+_D\cdot J^-_{D'} + J^-_D \cdot J^+_{D'}) }{1+\delta_{DD'}} .
\end{eqnarray}
 Precision data at and below the $Z$-peak are affected only by
$\mathcal{O}'_{LL}$~\cite{BS}. To study also LEP2 data above the
$Z$-peak, \cite{skiba} added 10 more 4-fermion operators. The full
list of the 11 4-fermion operators involving leptons is then given
by:
$$\mathcal{O}_{EE},
\mathcal{O}_{LL},\mathcal{O}_{EL},
\mathcal{O}_{EU},\mathcal{O}_{ED},\mathcal{O}_{EQ},
\mathcal{O}_{LU},\mathcal{O}_{LD},\mathcal{O}_{LQ},
\mathcal{O}'_{LL},\mathcal{O}'_{LQ}$$
\end{itemize}
In total this makes 18 operators. We here also consider 4 more oblique operators
(i.e.\  operators that do not involve fermions):
\begin{eqnsystem}{sys:O0}
\mathcal{O}_{WB} &=& \left( H^\dagger \tau^a H \right)\, W^a_{\mu \nu} B^{\mu \nu}
 = -2 \frac{M_W^2}{g^2}\, W^3_{\mu \nu} B^{\mu \nu} +\cdots\,;\\
\mathcal{O}_{HH} &=& |J_{\mu H}|^2
 =  \left( \frac{M_W^2}{g^2} \right)^2 (g W^3_\mu - g' B_\mu)^2 +\cdots\,;\\
 \mathcal{O}_{WW} &=&  \frac{(\mathcal{D}_\rho W^a_{\mu\nu})^2}{2}\\
 \mathcal{O}_{BB} &=& \frac{(\mathcal{D}_\rho B_{\mu\nu})^2}{2}
\end{eqnsystem}
Using the equations of motion for
the two neutral gauge bosons (see~\cite{GST} for a recent discussion),
these oblique operators can be reduced to combinations of the previous 18,
up to poorly constrained operators, e.g.\ operators that affect
couplings among vectors:
\beq
 i B_{\mu\nu} \mathcal{D}^\mu H^\dagger \mathcal{D}^\nu H,\qquad
 i W^a_{\mu\nu}\mathcal{D}^\mu H^\dagger \tau^a \mathcal{D}^\nu
H.
\eeq
This operator basis can be converted into the basis discussed in the main text
by using the equations of motion for
the gauge bosons $W^{\pm},W^3,B$ to eliminate all the
charged lepton currents from  $\Lag_{\rm BSM}$. At leading order in the
operator coefficients, we only need the equations of motion  that follow from $\Lag_{\rm SM}$:
\begin{eqnsystem}{sys:eqs}
\partial^\nu B_{\nu \mu} + \frac{M_W^2}{g^2} g' (g' B_\mu - g W^3_\mu) + g' \sum_F Y_F J^f_{\mu} & =&  0 +\ldots \\
\partial^\nu W^3_{\nu \mu} + \frac{M_W^2}{g^2} g (g W^3_\mu - g' B_\mu) + g \sum_f T_3 J^f_\mu & =&  0 +\ldots \\
\partial^\nu W^{\pm}_{\nu \mu} + M_W^2 W^+_\mu + \frac{g}{\sqrt{2}} \sum_F J^{\pm}_{\mu F} & =&  0 + \ldots
\end{eqnsystem}
where we neglected on the r.h.s.\ operators that are poorly
measured. We now  solve the equations of motion in terms of
\begin{equation}
J^{e_R}_\mu \equiv J_{\mu E}= \bar e_R \gamma_\mu e_R ,\qquad
J^{e_L}_\mu\equiv \bar e_L \gamma_\mu e_L ,\qquad
J^{+}_{L\mu} \equiv \bar e_L \gamma_\mu \nu_L
\end{equation}
and plug the result into the Lagrangian generated by the new
physics. In this way we replace $\Lag_{\rm BSM}$ with an equivalent
version that does not contain corrections to $Z e^+ e^-$ and $W^+ e^- \nu$ vertices
nor 4-fermion operators involving charged leptons.  The effects of
new physics have been completely recast on the propagators of the
gauge bosons and in the couplings of the gauge bosons to quarks and
neutrinos, and we obtain a Lagrangian in the form of eq.~(\ref{eq:lagrour}).
Thus, we can read off the parameters of
section~\ref{sec:parameters} in terms of the coefficients of the operators we listed above.

\subsection{Oblique corrections and neutrinos}

For
the oblique parameters we find:

\begin{eqnsystem}{sys:obliqueops}
\hat S &=& \frac{M_W^2}{g g'}\bigg[4c_{WB} + 4 t\, (c'_{HL}- c'_{LL}) + \frac{2}{t}\, (c_{HE}-c_{EE}- c_{EL} )
+ \nonumber \\
  && +2t\, (c_{HE} + 2 c_{HL} - c_{EE} - 2 c_{LL} - 3 c_{EL}) \bigg] \,,\\
\hat T &=&\frac{M_W^2}{g^2}\bigg[-2c_{HH} + 8 (c_{HL} + c_{HE}) - 4 (c_{LL} + c_{EE} + 2 c_{EL}) \bigg] \,,\\
\hat U & = & \frac{M_W^2}{g^2} \bigg[ 4 (c_{HE} + 2 c_{HL} ) - 4 (c_{EE} + 2 c_{LL} + 3 c_{EL})\bigg] \,,\\
V &=&  -\frac{M_W^2}{g^2} \bigg[c_{EE} + 4 c_{LL} + 4 c_{EL} \bigg] \,,\\
X & = & \frac{M_W^2}{g g'} \bigg[ c_{EE} + 2 c_{EL} \bigg] \, ,   \\
Y & = &  \frac{M_W^2}{g'^2} \bigg[2c_{BB} {g'}^2 - c_{EE} \bigg]   \,,\\
W &=& \frac{M_W^2}{g^2} \bigg[ 2c_{WW}g^2 -4 c'_{LL} - (c_{EE} + 4 c_{LL} + 4 c_{EL}) \bigg].
\end{eqnsystem}
Next, we give the expressions for the non-oblique terms defined in eq.~(\ref{eq:vert}).
The correction to the on-shell neutrino/$Z$ couplings is

\begin{eqnarray}
\delta g_{L\nu} & = & - \frac{2 M_W^2}{g^2} \Big[ c_{HE} +2 c_{HL} \Big]\, = V - \frac{1}{2} \hat U - t X \,.
\end{eqnarray}
We see that it can be re-expressed in terms of the oblique parameters.
This is true also for the corrections to the off-shell couplings
$C^\gamma_{L\nu}$ and $C^Z_{L \nu}$: we do not give their explicit expressions because
experiments negligibly constrain them.
This shows that the 7 oblique parameters fully describe charged leptons {\em and} neutrinos.

\subsection{Vertex corrections}

As discussed in the text, in the quark sector our approximation includes only
two more important combinations of effects. They are:

\begin{eqnarray}
\delta \varepsilon_q & = & \frac{2 M_W^2}{g^2} \Big[ 2 ( c'_{HQ} - c'_{HL} ) - (c_{HE} + 2 c_{HL} ) \Big]\, = \\
 & = &  \frac{4 M_W^2}{g^2}  ( c'_{HQ} - c'_{HL} ) + V - \frac{1}{2} \hat U - t X\,, \nonumber \\
\delta C_q & = & \frac{2 M_W^2}{g^2+{g'}^2} \Big[ 2 ( c'_{LQ}-c'_{LL}) - (  c_{EE} + 3  c_{LE} + 2 c_{LL})  \Big] \, =\\
& = & \frac{4 M_W^2}{g^2+{g'}^2} ( c'_{LQ}-c'_{LL}) + c^2 V - c s X\,. \nonumber
\end{eqnarray}
Besides oblique terms, they only depend on the operators involving SU(2)$_L$ currents.
The oblique approximation, therefore, fails only in the SU(2) sector.

For completeness, we also list the other parameters in the quark sector that we neglect: the 6 parameters involving left-handed quarks are
\begin{eqnarray}
\delta g_{Lq} & = & -2 \frac{M_W^2}{g^2} \Big[ (c_{HQ}-Y_Q c_{HE}) \pm ( c'_{HL} - c'_{HQ}) \pm \frac{1}{2} \,( c_{HE} + 2 c_{HL} ) \Big]\,, \label{eq:deltagLq}\\
C^\gamma_{Lq} & = & \frac{M_W^2}{e^2} \Big[ (c_{EQ}-Y_Q c_{EE}) (c^2-s^2) - 2 (c_{LQ} - Y_Q c_{LE} ) s^2  \pm 2 (c'_{LQ}-c'_{LL}) s^2  +\nonumber \\
&& \pm \frac{c^2}{2} (c_{EE} + 2 c_{EL} ) \mp \frac{s^2}{2} (c_{EE} + 4 c_{LL} + 4 c_{EL} ) \Big]\,,\\
C^Z_{Lq} & = & -\frac{2 M_W^2}{g^2+{g'}^2} \Big[ (c_{EQ}+c_{LQ}) - Y_Q (c_{EE} + c_{LE} ) \pm  ( c'_{LL}-c'_{LQ}) +\nonumber\\  & &\pm \frac{1}{2} ( c_{EE} + 3  c_{LE} + 2 c_{LL})  \Big]\,,
\end{eqnarray}
where $q$ stands for $u$ and $d$ and the signs refer to the up/down component of the doublet.
They depend on 5 coefficients $c_{HQ}$, $c'_{HQ}$, $c_{EQ}$, $c_{LQ}$, $c'_{LQ}$:
only the differences $C^Z_{Lu} - C^Z_{Ld}$ and $C^\gamma_{Lu} - C^\gamma_{Ld}$ depend on $c'_{LQ}$, and are related by
\begin{equation}
(C^\gamma_{uL}-C^\gamma_{dL}) = c^2 (C^Z_{u L}-C^Z_{d L}) - \frac{1}{t} X\,.
\label{relations2}
\end{equation}
In practice, we neglect $C^Z_{Lu} + C^Z_{Ld}$ and $C^\gamma_{Lu} + C^\gamma_{Ld}$, and $C^\gamma_{Lu} - C^\gamma_{Ld}$ is determined in terms of $\delta C_q$ and oblique parameters.

The corrections to the right-handed quark couplings are described by the following 6 parameters:
\begin{eqnarray}
\delta g_{Rq} & = & -2 \frac{M_W^2}{g^2} \Big[ c_{Hq} - Y_q c_{HE} \Big]\,,\\
C^\gamma_{Rq} & = & \frac{M_W^2}{e^2} \Big[ (c_{Eq} - Y_q c_{EE} ) (c^2- s^2) -2 (c_{Lq} - Y_q c_{EL}) s^2 \Big]\,,\\
C^Z_{Rq} & = & -\frac{2 M_W^2}{g^2+{g'}^2} \Big[ c_{Eq} + c_{Lq} - Y_q (c_{EE} + c_{EL}) \Big]\,,
\end{eqnarray}
where $q$ stands for $U$ and $D$.
They depend on the 6 coefficients $c_{HU}$, $c_{HD}$, $c_{EU}$, $c_{ED}$, $c_{LU}$, $c_{LD}$,
and are independent.
Their effect on precisio measurements is negligible.

Corrections to the quark/$W$ couplings are determined in terms of our parameters:
we do not give explicit expressions as experiments negligibly constrain these couplings.

For the bottom, $\delta \varepsilon_b$ can be read off from eq.~(\ref{eq:deltagLq}):

\beq
\delta \varepsilon_b  =  \frac{M_W^2}{g^2} \Big[ ( c'_{HQ_3} - c'_{HL}) + c_{HQ_3} - c_{HL} - \frac{2}{3} c_{HE} \Big]\,.
\eeq
In the flavour universal limit, it can be written in terms of the light quark parameters in the following way:

\beq
\delta \varepsilon_b &=& \delta \varepsilon_q  - (\delta g_{uL} + \delta g_{dL} )\,.
\eeq

\subsection{The universal limit}

One can verify that in the limit of heavy universal
new physics, our expressions reduce to the $\hat{S},\hat{T},W,Y$
parameters only, with all other parameters vanishing. Indeed in the
`universal' case only the following combinations of currents can
appear in $\Lag_{\rm BSM}$: \beq J_Y^\mu = \sum_f Y_F J_F^\mu,
\;\;\;\;\; J^a = \sum_D J_D^{\mu,a}. \eeq This restricts the
coefficients of the operators to be of the form \beq c_{HF} = Y_F\,
c_v\,, \quad c'_{HF} = c'_v\,, \quad c_{FF'} = Y_F Y_{F'}\, c_{4f}
\,, \quad c'_{FF'} = c'_{4f}\, , \eeq
such that the non-vanishing  $\hat S,\hat T,W,Y$ parameters are
\begin{eqnsystem}{sys:STWY}
\hat S &=& \frac{M_W^2}{g g'}\bigg[4 c_{WB} + 4 t\, (c'_{v} - c'_{4f}) + \frac{2}{t}\, c_v - \frac{1}{t}\, c_{4f} \bigg] \,,\\
\hat T &=&\frac{M_W^2}{g^2}\bigg[-2 c_{HH} + 4 c_v + c_{4f} \bigg] \,,\\
Y & = &  \frac{M_W^2}{g'^2} \bigg[2 c_{BB} {g'}^2 - c_{4f} \bigg]   \,,\\
W &=& \frac{M_W^2}{g^2} \bigg[ 2 c_{WW}g^2 - c'_{4f} \bigg].
\end{eqnsystem}
These expressions explicitly show how the 4 oblique operators in eq.~(\ref{sys:O0})
are equivalent to appropriate universal combinations of non-oblique operators.

\small
\begin{multicols}{2}

\end{multicols}
\end{document}